\documentclass[aps,prb,twocolumn,superscriptaddress,showpacs]{revtex4}

\usepackage{graphicx}

\begin{document}
\newcommand{\ket}[1]{\left|#1\right\rangle}
\newcommand{\bra}[1]{\left\langle#1\right|}
\newcommand{\braket}[2]{\left\langle#1|#2\right\rangle}
\newcommand{\im}{i}
\newcommand{\imag}{\grave{\imath}}

\title{Unbiased bases (Hadamards) for 6-level systems: Four ways from Fourier}

\author{A. J. Skinner}
\author{V. A. Newell}
\author{R. Sanchez}
\affiliation{Department of Physics, Skidmore College, Saratoga Springs, NY 12866}


\date{\today}

\begin{abstract}
In quantum mechanics some properties are maximally incompatible, such as the position and momentum of a particle or the vertical and horizontal projections of a 2-level spin.  Given any definite state of one property the other property is completely random, or unbiased.  For $N$-level systems, the $6$-level ones are the smallest for which a tomographically efficient set of $N+1$ mutually unbiased bases (MUBs) has not been found.  To facilitate the search, we numerically extend the classification of unbiased bases, or Hadamards, by incrementally adjusting relative phases in a standard basis.  We consider the non-unitarity caused by small adjustments with a second order Taylor expansion, and choose incremental steps within the $4$-dimensional nullspace of the curvature.  In this way we prescribe a numerical integration of a $4$-parameter set of Hadamards of order $6$.
\end{abstract}


\maketitle

\section{Introduction}
Mutually unbiased bases (MUBs) play an important role in quantum physics.  Typically they are constructed from the eigenstates of maximally incompatible properties.\cite{Schwinger}  For example, in cases ranging from quantum optics to electronics, the state $\rho$ of a photon polarization or electron spin is often defined by a Stokes or Bloch vector $\langle\vec{\sigma}\rangle$ whose components are the expectation values of the Pauli operators: $\rho = (I + \langle\vec{\sigma}\rangle\cdot\vec{\sigma})/2$.  The Pauli operators $\sigma_x$, $\sigma_y$, $\sigma_z$ are incompatible because, given an eigenstate of any one of them, the outcomes of the others are completely random.  Their eigenbases are said to be mutually unbiased.  In general, two bases $\ket{v_m}$ and $\ket{w_n}$ are unbiased when all probabilities between them are equal:
\begin{equation}
|\!\braket{v_m}{w_n}\!|^2 = 1/N \mbox{ for all } m,n
\end{equation}
with $N$ as a normalization constant.

MUBs are also manifest in Heisenberg's Uncertainty Principle: a definite momentum $\ket{p}$ implies a random position $\ket{x}$, corresponding to its wavefunction of equi-modular amplitudes
\begin{equation}
\Psi_p(x) = \braket{x}{p} =\frac{1}{\sqrt{2 \pi \hbar}} e^{\im p x/\hbar}.
\end{equation}

For a discrete system having $N$ orthogonal states, or levels, a set of $N+1$ MUBs gives a sufficient \cite{Ivanovic} and optimal \cite{WoottersFields} way to determine its possibly-mixed state $\rho$.  MUBs are thus useful for quantum state tomography.  They also play a role in quantum cryptography, \cite{Gisin} operator algebras, \cite{Popa} and Lie algebras.\cite{Kostrikin}  For a $2$-level system, or qubit, the Pauli operators provide a useful set of $3$ MUBs.  More generally, sets of $N+1$ MUBS have been constructed for systems whose number of levels $N$ is a prime number \cite{Ivanovic} or a power of a prime number. \cite{WoottersFields} In any case it is never possible to have more than $N+1$ MUBs.\cite{Delsarte}

For a $6$-level system, such as a spin-$5/2$ or a qubit-qutrit pair, there is a widespread and ongoing search for sets of MUBs.\cite{Bengtsson}  $6$ is the smallest integer that is neither a prime nor a power of a prime, and, despite the qualitative and physical appeal of sets of $N+1$ MUBS, no more than $3$ MUBs have been found \cite{Zauner,Butterley,Brierley} for the $6$-level systems.

Most searches begin with the classification of unbiased bases represented by Hadamards \cite{Hadamard} $U$ that are unitary and comprise, in a standard basis $\ket{s_i}$, equi-modular elements:
\begin{equation}
U^{\dagger} U  = I \mbox{ and } \bra{s_i}\!U\!\ket{s_m} = \frac{1}{\sqrt{N}} e^{\im \phi_{im}}.  
\end{equation}
The columns of $\bra{s_i}\!U\!\ket{s_m}$ form an orthonormal basis of vectors $\ket{u_m} \equiv U\ket{s_m}$ unbiased with respect to the standard basis; we are using the term ``unbiased basis'' relative to a standard basis.  Quantum mechanics allows a choice of overall phase for each $\ket{s_i}$ and $\ket{u_m}$ which we use to fix, in the first row and column, $\phi_{1m}=\phi_{i1}=0$.  Any reordering of the standard and unbiased bases' vectors (the permuting of rows and columns in $U$) is also immaterial.  

The classification of these equivalent Hadamards is an open question for $N>5$. \cite{Tadej}  For $N=6$ the known families of Hadamards have up to $2$ parameters and are topologically connected; \cite{Matolcsi} it has been conjectured there are $4$-dimensional families\cite{Bengtsson} which would vastly expand the searchable candidates for sets of MUBs.  The conjecture stems from an upper bound, know as the defect, on the dimensionality of any analytic set of Hadamards. \cite{Tadej}

Here we support the conjecture by numerical integration of Hadamards of order $6$ in $4$ directions within the $25$-dimensional space of relative phases $\phi_{im}$ ($i\ne1\ne m$).  We do this by Taylor-expanding to second order the non-unitarity of $6\times6$ matrices of equi-modular elements around known- and found-Hadamards.  Neighboring Hadamards lie nearby in the paraboloid valley of the expansion.  Apart from Tao's matrix, \cite{Tao} we always find the valley, i.e.\ the nullspace of the curvature, to be $4$-dimensional, consistent with the defect, although that nullspace can vary from Hadamard to Hadamard.  By taking small steps constrained to the changing nullspace, we numerically integrate Hadamards according to $4$ distinct parameters.

\section{Fourier and Other Known Families}
A common Hadamard is the Fourier matrix
\begin{equation}
\bra{s_j}F\ket{s_m} = \frac{1}{\sqrt{N}}e^{2 \pi \im (j-1) (m-1)/N}
\end{equation}
from which there originate two $2$-parameter affine families.  The first Fourier family, $F_6$, has components
\begin{equation}
F_6(\phi_1,\phi_2) = \frac{1}{\sqrt{N}}\left(
\begin{array}{cccccc}
1 & 1            & 1            & 1     & 1           & 1 \\
1 & q z_1     & q^2 z_2 & q^3 & q^4 z_1 & q^5 z_2 \\
1 & q^2        & q^4        & 1     & q^2        & q^4 \\
1 & q^3 z_1 & z_2        & q^3 & z_1        & q^3 z_2 \\
1 & q^4        & q^2        & 1     & q^4        & q^2 \\
1 & q^5 z_1 & q^4 z_2 & q^3 & q^2 z_1  & q z_2 \\
\end{array}\right)
\end{equation} 
with $q \equiv e^{2 \pi \im/N}$ and with $z_1 = e^{\im \phi_1}$ and $z_2 = e^{\im \phi_2}$ setting certain phases according to the affine parameters $\phi_1$  and $\phi_2$.  It is Hadamard because it comprises equi-modular elements and is always unitary --- its columns are an unbiased basis (with respect to the standard basis).  Its transpose $F_6^T(\phi_3,\phi_4)$, with $\phi_3$ and $\phi_4$ taking the (transposed) places of $\phi_1$ and $\phi_2$, is the second Fourier family.

There are other Hadamard families: Beauchamp's and Nicoara's $B_6(y)$ \cite{Beauchamp} which interpolates from Bj\"{o}rck's circulant matrix $C_6$ \cite{Bjorck} and its conjugate to Di\c{t}\u{a}'s $D_6(\theta)$; \cite{Dita} and Matolcsi's and Sz\"{o}ll\H{o}si's $M_6(x)$ which connects $D_6$ with $F_6$ and $F_6^T$.\cite{Matolcsi}  There is also Tao's matrix $S_6$ comprising $3$rd roots of unity, \cite{Tao} which is disconnected from the rest. \cite{Bengtsson}

\section{Non-unitarity to second order}
All these Hadamards are drawn from the matrices of equi-modular elements $\bra{s_i}E(\vec{\phi})\ket{s_m} = e^{\im \phi_{im}}/\sqrt{N}$ which we take to be a function on a $25$-dimensional phase space, with $\vec{\phi}$ having the individual phases as components:
\begin{equation}
\vec{\phi} = (\underbrace{\phi_{22},\ldots,\phi_{62}}_{\mbox{2nd column}},\underbrace{\phi_{23},\dots,\phi_{63}}_{\mbox{3rd column}},\ldots,\underbrace{\phi_{26},\ldots\phi_{66}}_{\mbox{6th column}}),
\end{equation} 
excluding the $\phi_{1m}=\phi_{i1}=0$ in the matrices' first row and column.

The difficulty is in choosing the phases to ensure a matrix is unitary.   We therefore define a measure of its non-unitarity as proportional to the sum of the probabilities between non-trivial pairings of its column vectors,  
\begin{equation}
f \equiv (N/2) \sum_{n>m} |\langle e_m|e_n\rangle|^2 \ge 0,
\end{equation}
which vanishes only when $E$ is unitary (and thus a Hadamard).  Taking the inner product $\braket{e_m}{e_n}$ and its conjugate in the standard basis, we find the non-unitarity $f$ is a simple function of the relative phases:
\begin{equation}
f=\sum_{n>m}\sum_{j>i}\cos(\phi_{in}-\phi_{im}+\phi_{jm}-\phi_{jn})+\sum_{n>m}\sum_{j=i}\frac{1}{2}.
\end{equation}
\vspace{0.01in}

Our methods will rely on Taylor-expanding the non-unitarity to second order in small phase shifts $\Delta\vec{\phi}$ as
\begin{equation}
f(\vec{\phi}+\Delta\vec{\phi}) = f(\vec{\phi}) + \vec{g}\cdot \Delta\vec{\phi} + \frac{1}{2}\Delta\vec{\phi}\cdot H \cdot \Delta\vec{\phi} + {\cal{O}}(\Delta\vec{\phi}^3),
\end{equation}
with $\vec{g}$ as the gradient (slope) and $H$ as the Hessian (curvature) of the non-unitarity.

The gradient $\vec{g}$ has $25$ components 
\begin{equation}
\frac{\partial f}{\partial \phi_{ko}} = -\sum_{n>m}(\delta_{no}-\delta_{mo}) \sum_{j>i}(\delta_{ik}-\delta_{jk})\sin(\star)
\end{equation}
with $\sin(\star)$ taking the same argument as before: $\star = \phi_{in}-\phi_{im}+\phi_{jm}-\phi_{jn}$.  It is subject to the chain rule for differentiation and thus gives rise to  the Kronecker deltas.  The gradient also vanishes for Hadamards because they minimize the non-unitarity.

The $25\times25$ components of the Hessian $H$ are
\begin{widetext}
\begin{equation}
\frac{\partial^2 f}{\partial \phi_{ko} \partial \phi_{lp}} = -\sum_{n>m} (\delta_{no}-\delta_{mo}) (\delta_{np}-\delta_{mp}) \sum_{j>i} (\delta_{ik}-\delta_{jk}) (\delta_{il}-\delta_{jl})\cos(\star).
\end{equation}
\end{widetext}
The Hessian is a real symmetic matrix whose eigenvalues give the curvature of $f(\vec{\phi})$ in the directions of their corresponding eigenvectors (principal axes).  

Whenever $\vec{\phi}$ points to a Hadamard, both $f$ and $\vec{g}$ are zero.  Moving away by a small step $\Delta\vec{\phi}$ changes the gradient according to its first order Taylor expansion,
\begin{equation}
\vec{g}(\vec{\phi}+\Delta\vec{\phi}) =\vec{g}(\vec{\phi}) + H(\vec{\phi}) \cdot \Delta\vec{\phi} + {\cal{O}}(\Delta\vec{\phi}^2).
\end{equation}
To preserve unitarity we require no change to $\vec{g}=0$ (so that $\vec{\phi} \Rightarrow \vec{\phi}+\Delta\vec{\phi}$ continues to minimize the non-unitarity).  We are therefore primarily interested in the nullspace of $H$ (those vectors $\vec{n}$ for which $H\cdot \vec{n}=0$).  

\section{Numerical Methods and Results}
We calculate numerically \cite{Mathematica} the non-unitarity $f$, gradient $\vec{g}$, and Hessian $H$ as functions of $\vec{\phi}$; from $H$ we calculate its eigenvalues $\lambda_i$ and eigenvectors $\hat{n}_i$.  When $\vec{\phi}$ points to a Hadamard, with the exception of Tao's Matrix $S_6$, we always find a $4$-dimensional nullspace: four eigenvectors $\hat{n}_i$ point in directions of effectively-zero curvature, i.e.\ they have numerical eigenvalues $\lambda_i\le10^{-4}$.  We keep the four eigenvectors $\hat{n}_i$ that span the nullspace and discard the rest.

\begin{figure}
\begin{center}
\includegraphics{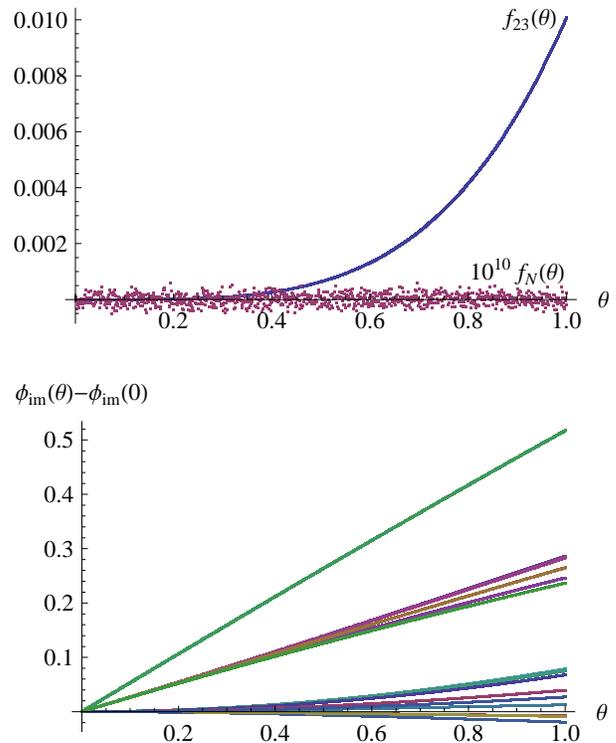}
\end{center}
\caption{Non-unitarity $f_{23}(\theta)$ in the $\hat{\phi}_2+\hat{\phi}_3$ direction, from Fourier; small stepping within the nullspaces improves unitarity by $10$ orders of magnitude (see scatter plot $10^{10} f_N(\theta)$).  Lower panel: the numerically integrated phase shifts.  Staying within nullspaces causes a spreading of the phases from their original rates of change $0$, $1/\sqrt{12}$, and $1/\sqrt{3}$.}
\label{evolved23}
\end{figure}

From the Fourier matrix $F$ we already knew of four directions to move away, namely the $\hat{\phi}_i$ corresponding to the directions of increasing the parameters $\phi_i$ of the two $2$-parameter Fourier families, e.g.\,
\begin{equation}
\hat{\phi}_1 \equiv (\underbrace{1,0,1,0,1}_{\mbox{2nd column}},0,0,\ldots,0,\underbrace{1,0,1,0,1}_{\mbox{5th column}},0,0,\ldots,0)/\sqrt{6}.
\end{equation}
Although the $\hat{\phi}_i$ span the nullspace, they are not all mutually orthogonal or compatible between the two families; moving a finite distance in the $\hat{\phi}_2+\hat{\phi}_3$ direction, for example, quickly loses the requisite unitarity; see Figure \ref{evolved23}.

The existence of the nullspaces, however, suggests that a series of infinitesimal moves might preserve unitarity. The problem is that the nullspace changes from Hadamard to Hadamard so that a direction constrained to the nullspace at one point might depart from the nullspace at another.  We propose to ``feel a way forward'' along an evolving direction $\vec{\theta}$ obtained by projecting the previous step's direction into the new nullspace:
\begin{equation}
\vec{\theta}_{[c]} \equiv \sum_{i=1}^4 \hat{n}_{i[c]}\;\hat{n}_{i[c]} \cdot \vec{\theta}_{[c-1]},
\end{equation}
with $c$ as a step index.  Each step adjusts the relative phases $\vec{\phi}$ by a length $\Delta \theta$ in that direction:
\begin{equation}
\vec{\phi}_{[c+1]} = \vec{\phi}_{[c]} + \Delta \theta \; \hat{\theta}_{[c]}.
\end{equation}
At each step we also use Newton's method to correct for any higher order drift $({\cal{O}}(\Delta\theta^2))$ off the $\vec{g}=0$ valley floor:
\begin{equation}
\vec{\phi}_{[c+1]} \Rightarrow \vec{\phi}_{[c+1]} - \tilde{g}_{[c+1]}/\tilde{H}_{[c+1]},
\end{equation}
with $\tilde{g}$ and $\tilde{H}$ as the projections of the gradient and Hessian outside the nullspace; at this point we recalculate the Hessian and its nullspace.  These corrective steps are small (never longer than $1\%$ of $\Delta\theta$) and allow us to take finite, rather than infinitesimal, small steps $\Delta\theta=0.001$ (radians).

In this way we numerically integrate a parameterized curve of Hadamards through the space of matrices of equi-modular elements.  Beginning with $\vec{\phi}$ specifying the Fourier matrix $F$, for example, and taking a small step in the $\hat{\phi}_2+\hat{\phi}_3$ direction, for example, we improve it with Newton's method, obtain a new direction from the projection of the previous, take another small step, improve it with Newton's method, and so on, progressing a distance $\theta$ while maintaining the unitarity, numerically discovering this curve of Hadamards originating from $F$ and the initial choice of direction; see Figure \ref{evolved23}.

As we integrate curves in this manner, we similarly evolve four directions $\vec{\theta}_{i}$ spanning the evolving nullspace, beginning with, in our example, the four $\hat{\phi}_i$ from the two Fourier families.  In any case, some initial choice of nullspace spanning vectors $\vec{\theta}_{i[1]}$ can be made.  Then with each small step along a curve we obtain new directions from the previous vectors' projections into the new nullspace,
\begin{equation}
\vec{\theta}_{i[c]} \equiv \sum_{j=1}^4 \hat{n}_{j[c]}\;\hat{n}_{j[c]} \cdot \vec{\theta}_{i[c-1]},
\end{equation}
just as we do for the curve's evolving direction $\vec{\theta}_{[c]}$.

The evolving four directions $\vec{\theta}_{i[c]}$ suggest a way to integrate Hadamards of order 6 according to four distinct parameters $\theta_i$: we numerically integrate a distance $\theta_1$ in the first evolving $\vec{\theta}_{1[c]}$ direction, then a distance $\theta_2$ in the second evolving direction, and so on, obtaining a Hadamard an integrated distance $\theta=\sum_i\theta_i$ along the curve of four joined segments.  

\begin{figure}
\begin{center}
\includegraphics{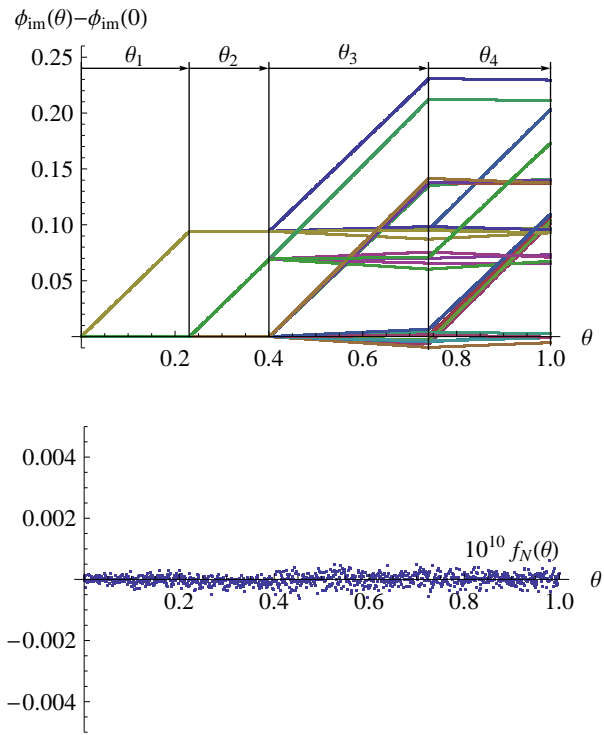}
\end{center}
\caption{Four-way foray from Fourier.  Upper panel: the numerically integrated phase shifts obtained by integrating a distance $\theta_1 = 0.23$ in the first evolving direction, $\vec{\theta}_{1[c]}$; then $\theta_2=0.17 $ in the second; $\theta_3=0.34$ in the third; and $\theta_4=0.26$ in the fourth.  The first two parameters correspond to moving within the first Fourier family; the last two parameters move away from it.  Lower panel: negligible non-unitarity $f_N(\theta)$ as a function of integrated distance $\theta$ along the four joined curves.} 
\label{evolved1234}
\end{figure}

In our example, beginning from the Fourier matrix and with the $\hat{\phi}_i$ giving the four initial directions, we obtain a $4$-parameter set of numerically integrated Hadamards
\begin{equation}
F_6(\theta_1,\theta_2,\theta_3,\theta_4).
\end{equation}
This is our main result.  The first two parameters specify Hadamards within the first Fourier family,
\begin{equation}
F_6(\theta_1,\theta_2,0,0)=F_6(\theta_1,\theta_2),
\end{equation}
and the second two parameters then specify Hadamards obtained by numerically integrating away from $F_6(\theta_1,\theta_2)$ and away from the first Fourier family in general; see Figure \ref{evolved1234}.

\section{Summary: Four Ways From Fourier.}
The $6$-level system is the smallest $N$-level system for which a tomographically complete and efficient set of $N+1$ MUBs has not been found; there is a widespread search for more than $3$.  A complete classification of the complex Hadamards of order $6$, representing bases unbiased with respect to a standard basis, would greatly narrow the search.  The previously known families of these Hadamards are topologically connected, perhaps by a conjectured $4$-parameter family.  

Here we prescribe the numerical integration of a $4$-parameter set of Hadamards within the space of matrices of equi-modular elements.  We Taylor expand to second order a measure of their non-unitarity and use the $4$-dimensional nullspace of the curvature to choose small steps that preserve the requisite unitarity of the Hadamards.  We also update, or evolve, a set of four directions, as we move along a curve of Hadamards, by projecting them into the evolving nullspace.

From the Fourier matrix $F$ we choose four initial directions (two from the first Fourier family and two from its transpose) and numerically integrate a $4$-parameter set of Hadamards $F_6(\theta_1,\theta_2,\theta_3,\theta_4)$ by integrating, in order of increasing $i$, distances $\theta_i$ in their corresponding evolving directions $\vec{\theta}_{i[c]}$.

These results should facilitate numerical searches for at least $4>3$ MUBs in only $12$ parameters ($4$ for each candidate MUB beyond the standard basis).  They might also help point the way to the complete analytical classification of Hadamards for $N=6$.  In that case, then, the MUB problem for $6$-level systems might finally be resolved.

{\bf Acknowledgment} This work is supported in part by a grant from
the Skidmore College Faculty/Student Summer Research Program.

\end{document}